\documentclass{article}

\usepackage{arxiv}

\usepackage[utf8]{inputenc} 
\usepackage[T1]{fontenc}    
\usepackage{hyperref}       
\usepackage{url}            
\usepackage{booktabs}       
\usepackage{amsfonts}       
\usepackage{nicefrac}       
\usepackage{microtype}      
\usepackage{lipsum}
\usepackage{cite}
\usepackage{amsmath,amssymb,amsfonts}
\usepackage{algorithmic}
\usepackage{graphicx}
\usepackage{textcomp}
\usepackage{xcolor}
\usepackage[font=small, labelfont=bf]{caption}
\title{Defined the predictors of the  lightning  over India by using artificial neural   network}

\author{
    Pradip Kumar Gautam\thanks{Use footnote for providing further
    information about author (webpage, alternative
    address)---\emph{not} for acknowledging funding agencies.} \\
    School of earth ocean and atmosphere sceinces\\
    Indian Institude of Technology \\
    Bhubaneswar, Odisha, India\\
   \texttt{pkg11@iitbbs.ac.in} \\
   \And
Deweshvar Singh \\
  School of earth ocean and atmosphere sceinces\\
  Indian Institude of Technology \\
  Bhubaneswar, Odisha, India\\
  \texttt{ds23@iitbbs.ac.in} \\
}

\begin{document}
\maketitle

\begin{abstract}
Lightning casualties cause tremendous loss to life and property. However, very lately lightning has been considered as one of the major natural calamities which is now studied or monitored with proper instrumentation. In this study, the lightning characteristics over India has been studies by using daily data LRTS (low resolution time saris) and monthly data HRMC (high resolution monthly climatology) of LIS/OTD during the period 2000-2013. LIS (lightning imaging sensor) provides smaller global coverage than OTD (Optical Transient Detector) and it is still thought to have detected 90 percent of the world’s lightning on annual basis. We have used ANN time series method ( a neural network ) to analyze the time series and defined which one will be best predictor of lightning over India. I have taken the the time series of lightning as output(dependent ) and input (independent ) as k-index, AOD, Cape, Relative humidity( 500 to 1000 pressure level) and Vertical integral of the divergence of cloud frozen of water flux and Vertical integral of divergence the cloud liquid of water flux.These input variable are analyzed by GPR, SVM, RT and LR show approximately  linear relation. 

\end{abstract}


.

\begin{section}{Introduction}
Natural disasters have been causing huge loss to life and property. However, lightning casualties have not been given due attention in the recent history. There are no doubt random studies and research going on, but it has not yet become one of the prime focus of the disaster response forces unlike cyclones, floods, forest fire, landslides etc. It has been recorded that 5259 people lost their lives due to lightning strikes in India during 1979 to 2011. Among the fatalities, 89 percent are male and are from the state of Maharashtra, West Bengal, Uttar Pradesh and others (Singh Singh, 2015). Lightning has been recognized as one of the most powerful, spectacular and all-pervasive atmospheric hazards that mankind has encountered throughout history (Cooray et al., 2007). Statistically it has been assessed that 1.4 billion flash rates (responsible for lightning) have been recorded annually (both inter-cloud and cloud to ground) over the entire Earth which translates into 44+/-5 lightning flashes every second around the globe (Christian et al., 2003). It has also been established that globally the frequency of lightning is the highest in the region between 30 N to 30 S. Thus, India after Africa can be regarded as one of the most vulnerable regions to lightning disasters.
\begin{figure}
\centering
\includegraphics[width=0.7\textwidth]{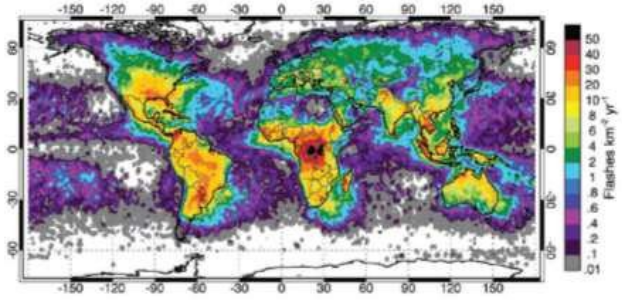}
\caption{The annualised distribution of total lightning activity (flash/per km/year) (Christian et al., 2003)}
    \includegraphics[width=0.7\textwidth]{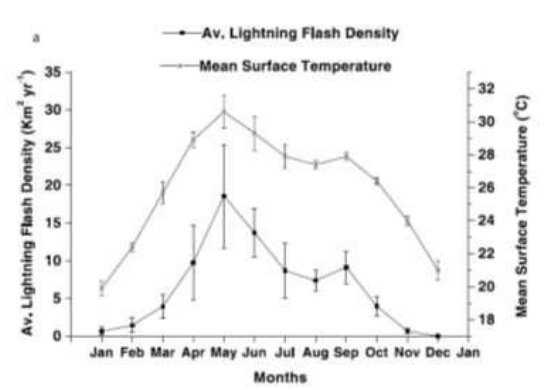}
\caption{Seasonality of Lightning and Mean Surface Temperature over India during 1998-2002 (Kandalgaonkar et al., 2005).}
\end{figure}
\end{section}
\section{Data and Methodology}
TRMM (Tropical Rainfall Measuring Mission) which have five different type of the sensor that is used for different purposes. It is flown on the Tropical Rainfall Measuring Mission (TRMM) satellite as part of the National Aeronautics and Space Administration’s (NASA) Earth Observing System (EOS). The LIS sensor was designed with a higher sensitivity and spatial accuracy than the OTD sensor. The LIS (Lightning Imaging Sensor and OTD (Optical Transient Detector) are two of them which are used for this study. The LIS/OTD gridded Lightning datasets consist of flash rate climatology, raw flashes, scaled flashes, flash rate, and flash rate time series data which have uniform flash detection efficiency, 93±4 \% in night and 73±11 \% day. The independent variables as k-index, AOD, Cape, Relative humidity( at 500 to 1000 pressure level) and Vertical integral of the divergence of cloud frozen of water flux and Vertical integral of divergence the cloud liquid of water flux datasets are taken from the era-5.
\begin{center}
\begin{tabular}{c|c|c|c}
\hline
Variable & Dimension & Unit & Resolution $^o$C\\
\hline
AOD & dimensionless & unitless  & 1x1 \\
\hline
Lightning & 5845x144x72 & count/$km^2$/day & 1x1 \\
\hline
cape & 5845x144x72 & j/kg & 1x1 \\
\hline
k-index & 5845x144x72 & c & 1x1 \\
\hline 
RH & 5845x144x72 & g/kg & 1x1 \\
\hline
Vertical velocity & 5845x144x72 & m/sec & 1x1 \\
\hline
\end{tabular}
\end{center}

\subsection{Artificial Neural Network}
The basic element of a FFNN is the neuron, which is a logical-mathematical model that seeks to simulate the behaviour and functions of a biological neuron. Figure 3 shows the schematic structure of a neuron. Typically, a neuron has more than one input. The elements in the input vector P= (P{1},P{2} ,...., P{R})are weighted by elements (w{1}, w{2},.........,w{j}) of the weight matrix W respectively.
\begin{figure}[!h]
    \centering
    \includegraphics[width=0.7\textwidth]{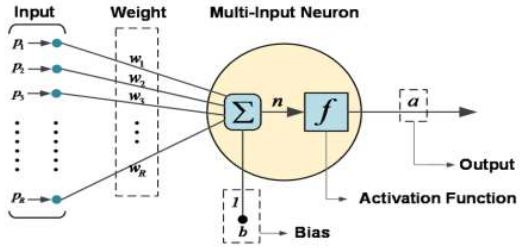}
    \caption{structure of the multi-input neuron.}
    \label{fig:bv}
\end{figure}
\subsection{Training Algorithm and Parameter}
\subsubsection{Training Parameter}
This algorithm typically requires more memory but less time. Training
automatically stops when generalization stops improving, as indicated by an
increase in the mean square error of the validation samples.
\begin{center}
\begin{tabular}{l|l}
   \hline
   Parameter   & Value\\
   \hline
   Epoch between displays & 2\\
   \hline
   Maximum epochs & 10\\
   \hline
   Time & second\\
   \hline
   Maximum Validation Failure & 0\\
   \hline
   Performance Goal & 0\\
   \hline
   Minimum Gradient error & $3.831e^{-08}$\\
   \hline
   Initial & $10^{-3}$\\
   \hline
   Maximum $\mu$ & $10^{10}$\\
   \hline
      No. of neurons used for hidden layer & 10\\
   \hline
\end{tabular}
\end{center}
\subsubsection{Activation Function}
In this study, the log-sigmoid activation function is adopted. It can be given by the
following expression:
\begin{equation}
 f(x)=1/(1+ e^{-x})   
\end{equation}
\subsection{Nonlinear auto-regressive model with exogenous inputs (NARX)}
The nonlinear auto-regressive network with exogenous inputs (NARX) is a recurrent
dynamic network, with feedback connections enclosing several layers of the network.
The NARX model is based on the linear ARX model, which is commonly used in time-
series modelling.
Predict series y (t) given d past values of y(t) and another series x (t).
\begin{equation}
y(t) = f(x(t-1),(t-2),...,x(t-d), y(t-1), y(t-2),....y(t-d))
\end{equation}

\subsection{Correlation Coefficient (r) }
The correlation coefficient indicates the strength and
direction of a linear relationship between two random
variables.
\begin{equation}
r = \frac{\sum_{i=1}^n((x{i}-\bar{x}) (y{i}-\bar{y})} {\sqrt{ \sum_{i=1}^n(x{i}-\bar{x})^2 (y{i}-\bar{y})^2}}
\end{equation}
The value of r is such that -1 < r < +1. The +ve and –ve signs are used for positive linear correlations and negative linear correlations, respectively. A correlation greater than 0.8 is generally described as strong, whereas a correlation less than 0.5 is generally described as weak.

\subsection{Mean Squared Error}
 Mean Squared Error is the average squared difference between outputs and
targets. Lower values are better and Zero means no error.

\begin{equation}
MSE = \frac {\sum_{i=1}^n(t{i}-o{i})^2} {p}
\end{equation}
where t{i} target value, o{i} output and p-pattern. ANN divides the time series data in to 70 percentage in training and 15 percentage in testing and 15 percentage in validation. I have done with 10 neurons (nodes) and with 2 hidden layer. After that ,I have trained the data by Levenberg-Marquardt method because it has much accuracy. The ANN gives the first result, how the each predictor perform with lightning which is given into the figure 5 . The performance is nothing but current status of
the training process. This graph is showing the current status of the training process. The X - axis indicates the number of iterations (5 Epochs). The Y- axis represents the MSE occurred for each iteration. The line graph plotted in blue colour represents the training results. The graph with green colour represents the validation. Results and the graph in red colour represent the test results. The performance graph is computed for every iteration in the training process and the graph in which all the three results of training, validation and testing coincide at almost all points is chosen to be the best performance. At that point of time the training should be stopped and no further iteration should be preceded. It means that no further training is required and if done, may predict the results wrongly.

\subsection{Root Mean Square Error}
The root mean square error (RMSE) is a frequently used measure of the differences between values predicted by a model and the values actually observed. It measures average error, weighted according to the square of the error. It does not indicate the direction of the deviation.
\begin{equation}
RMSE = \sqrt{\frac {\sum_{i=1}^n(t{i}-o{i})^2} {p}}
\end{equation}

\begin{section}{Result}
\begin{subsection} {Defined the variable which affects the lightning by LR, RT, GPR and SVM}
The regression learner app of the MATLAB software 2018a trains regression models to predict data. Using this app, we can explore your data, select features, specify validation schemes, train models, and assess results. we can perform automated training to search for the best regression model type, including linear regression models, regression trees, Gaussian process regression models, support vector machines, and ensembles of regression trees. Perform supervised machine learning by supplying a known set of observations of input data (predictors) and known response (lightning). Use the observations to train a model that generates predicted responses for new input data.

\begin{figure}
\centering
\includegraphics[width=1.0\textwidth]{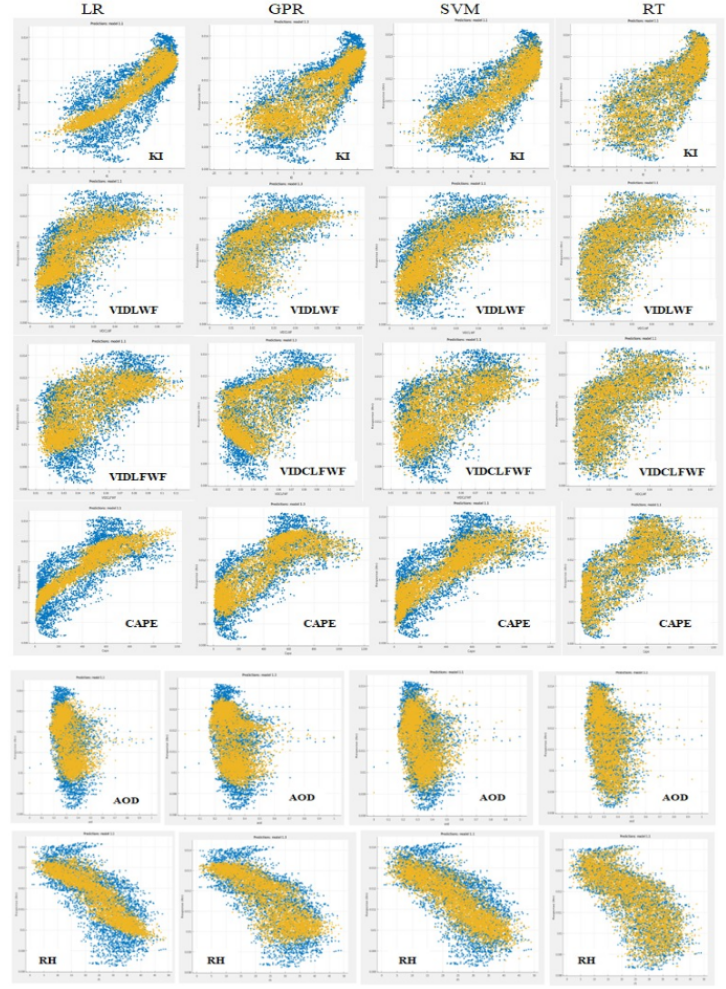}
\caption{This is the scatter plot of Lightning (dependent) and meteorological variables  by LR, GPR, SVM,  RT models. Only those variable are here, which are vary with lightning. }
\end{figure}
We have taken number of the meteorological data to establish the, best predictor of the lightning.
Firstly, The GPR model is showing 0.76 maximum value of R2 from other Model in Table. 1. That means the response variable lightning AOD, RH, Cape, K-Index, VIDCLWF and VIDCFWF are better correlated to the lightning. The GPR is not showing maximum R value but also it is representing the minimum RMSE (root mean square error), and maximum training time. Prediction speed is less but higher than support vector machine. Secondly, the regression trees model is representing the maximum R value. In last we have combined the all model and trained that one, we are getting that. It is not showing the better result than GPR and regression trees model. In figure 1 the blue point is the actual data and the yellow one is the predicted data. 

\begin{center}
\begin{tabular}{c|c|c|c|c|c}
\hline
Models & $R^2$ & R & RMSE & Prediction speed & Training Time\\
\hline
LR & 0.62 & 0.7874 & 0.00073867 & 260000 obs/sec & 3.4078sec\\
\hline 
SVM & 0.59 &0.7681 &0.00067807 & 5200 obs/sec & 19.967 sec\\ 
\hline
GPR & 0.76 & 0.8717 & 0.00070017  & 8300 obs/sec & 322sec\\
\hline
RT & 0.65& 0.8717& 0.00072151 & 360000 obs/sec & 1.4704 sec\\
\hline
All Models & 0.62 & 0.7873 & 0.00084561 & 71000 obs/sec & 11.768sec\\
\hline
\end{tabular}
\end{center}
In table 1, we can see that the RMSE ( Root mean Square Error)  and R(correlation coefficient) and R2(square of the correlation coefficient) of the variable is given. R-square is the percentage of the response variable variation that is explained by a linear model. R-square is always between 0 to 100 percentage. 0 percentage indicates that the model explains none of the variability of the response data around its mean.100 percentage indicates that the model explains all the variability of the response data around its mean. the higher the R-squared, the better the model fits our data. 
\end{subsection}
\begin{subsection}{experiment-1}
In given fig 5 we can see that RH (Relative Humidity), AOD ( aerosol optical depth) k-index, Cape (convective available Potential energy) VIDCFWF ( vertical Integral of the divergence of the cloud liquid frozen water flux) and VIDCLWF (vertical Integral of the divergence of the cloud liquid water flux) is showing best performance with lightning at 9 Epoch at which point the training , testing and validation are coincide.
\begin{figure}[!t]
    \centering
    \includegraphics[width=0.7\textwidth]{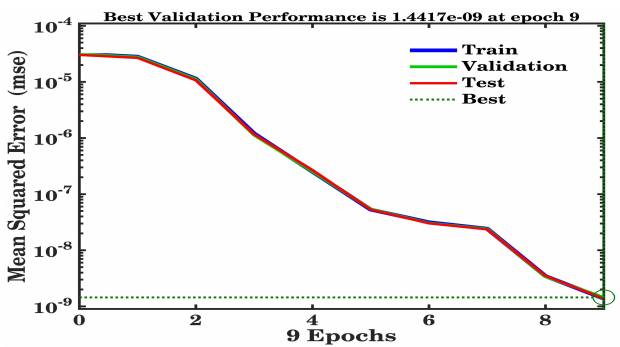}
    \caption{This is the performance of the predictor AOD, VIDCLFWF, KI, Cape, VIDCLWF, RH, with lightning during training.}
    \label{fig:d}
\end{figure}
\end{subsection}

\begin{subsection}{Experiment-2}
In figure 6, the x axis is the epoch (iteration) and y-axis is gradient descend, cross-entropy and validation fail.The figure 6 is showing the training state of the lightning and and independent variables in which we can tell about the gradient descend, cross entropy and validation. The performance (1447e-09) indicating how much minimised errors occur during the training, Gradient (3831e-08) indicating how much variance occurs in the error rate, Mu(cross entropy with 1e-10) is the threshold value for each iteration which is updated for each iteration and the Validation Check indicates whether the currently completed iteration (has minimized error compared to the previous, iterations.

\begin{figure}
    \centering
    \includegraphics[width=0.7\textwidth]{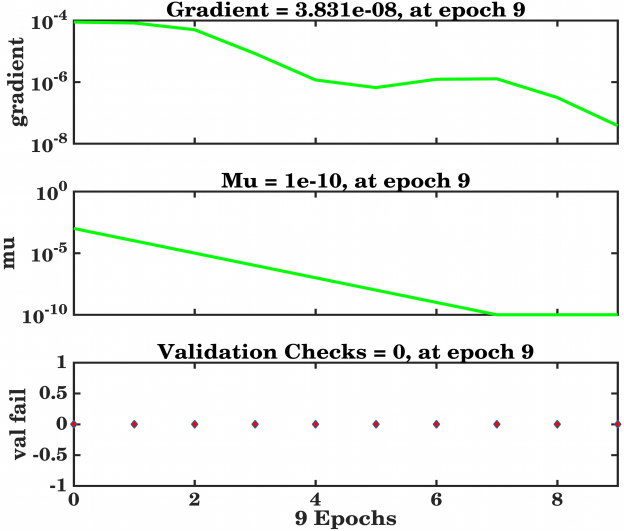}
    \caption{This is the performance of the predictor AOD, VIDCLFWF, KI, Cape, VIDCLWF, RH, with lightning during training.}
    \label{fig:f}
\end{figure}
\end{subsection}
\subsection{Experiment-3}
Figure 7 is the error histogram with 20 bin of the Training, validation and Test during training processes of the Model. In these histogram , There is a vertical yellow line in each which is represent the zero error . x- axis is the error and y-
axis is Instances( occurrence of number of the error) in each figure, each predictor’s have range of error of the training, validation and test.

\begin{figure}
    \centering
    \includegraphics[width=0.8\textwidth]{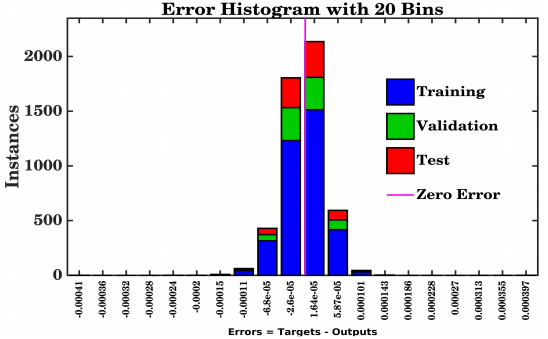}
    \caption{This is error histogram by which we can tell model are best or fail.}
    \label{fig:g}
\end{figure}

\begin{subsection}{experiment-4}

\begin{figure}
    \centering
    \includegraphics[width=0.7\textwidth]{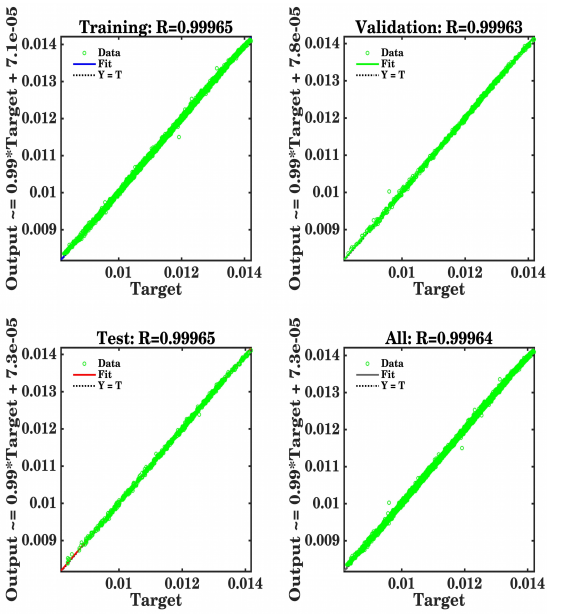}
    \caption{This figure is representing that model is well fitting or not.}
    \label{fig:h}
\end{figure}
This ( figure-8) is the regression plots to know the correlation during training,validation and Test, of lightning with each predictor’s. Regression R values measure the correlation between outputs and targets. An R value of 1 means a close relationship and MSE vale is also very less that's why fitting is well. Regression R values of 0 means a random relationship.
\end{subsection}
\begin{subsection}{experiment 5}
The Figure 9 describes how the prediction errors are related in time. For a perfect prediction model, there should only be one nonzero value of the auto correlation function, and it should occur at zero lag. (This is the mean square error.) This would mean that the prediction errors were completely uncorrelated with each other (white noise). If there was significant correlation in the prediction errors, then it should be possible to improve the prediction, perhaps by increasing the number of delays in the tapped delay lines. In this case, the correlations, except for the one at zero lag, fall approximately within
the 95percent confidence limits around zero, so the model seems to be adequate. If even more accurate results were required, you could retrain the network by clicking Retrain in ntstool. This will change the initial weights and biases of the network, and may produce an improved network after retraining.
\begin{figure}
    \centering
    \includegraphics[width=0.7\textwidth]{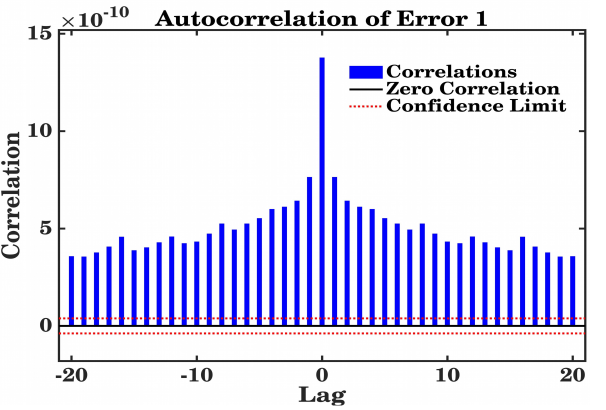}
    \caption{This is the auto correlation of error data which averaged as it divided by the number of samples in error data.}
    \label{fig:k}
\end{figure}
\end{subsection}

\begin{subsection}{experiment-6}
The figure 10 is showing correlation input and error 1. The error 1 is subtraction of the target and output.This input-error cross-correlation function illustrates how the errors are correlated with the input sequence x(t). For a perfect prediction model, all of the correlations should be zero. If the input is correlated with the error, then it should be possible to improve the prediction, perhaps by increasing the number of delays in the tapped delay lines. In this case, all of the correlations fall within the confidence bounds around zero.
\begin{figure}[!h]
    \centering
    \includegraphics[width=0.8\textwidth]{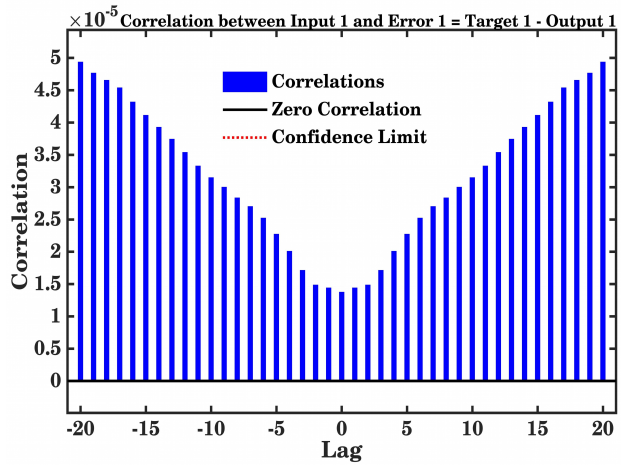}
    \caption{This is the auto correlation of error data which averaged as it divided by the number of samples in error data.}
    \label{fig:lm}
\end{figure}
\end{subsection}

\begin{subsection}{experiment-7}
This figure 11 displays the inputs, targets and errors versus time. It also indicates which
time points were selected for training, testing and validation.
The vertical yellow line is lie in between plus (blue color) mark and point (blue
color) mark in figure 6, that is representing error during training. Similarly
green and red color is indicating for validation and testing respectively.
\begin{figure}[!t]
    \centering
    \includegraphics[width=0.9\textwidth]{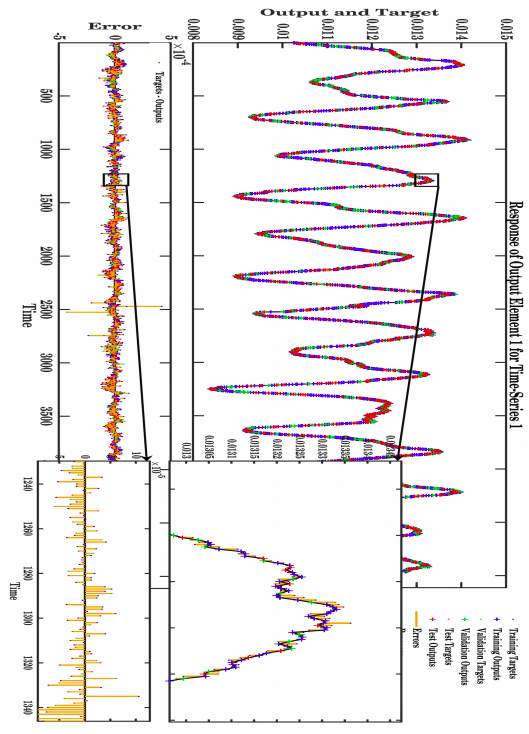}
    \caption{This is the time series response for the output element}
    \label{fig:ij}
\end{figure} 
\end{subsection}
\end{section}

\section{Conclusion}
In this experiment we have taken number of meteorological variables to defined the relative
variable of lightning by using four models LR, SVM, GPR, RT and combine of thses four model.
As a result cape, k-index, RH, AOD, vertical velocity, Vertical integral of the divergence of cloud
frozen of water flux and Vertical integral of divergence the cloud liquid of water flux are showing
good relation with lightning which are defined on the basis of the value of R and R 2 o f different
model. Finally artificial neural network are used to defined much better relation based on MSE and
regression values which are given in below table.The regression approximately one and mean
square error is very less. According to above the study,the MSE is very less as well as regression
value is nearly approximately one. So the result is that these variables named cape, k-index, RH,
AOD, vertical velocity, Vertical integral of the divergence of cloud frozen of water flux and Vertical
integral of divergence the cloud liquid of water flux, are good predictor of the lighting over India.

\begin{tabular}{c|c|c|c}
     \hline
     Data Type & Target Value & MSE & Regression\\
     \hline
     Training & 3581 & $1.37698e^{-9}$ & 0.99645\\
    \hline
     Validation & 767 & $1.44171e^{-9}$ & 0.99634\\
     \hline
     Testing & 767 & $1.36990e^{-9}$ & 0.99652\\
     \hline
\end{tabular}

According to above the study,the MSE is very less as  well as regression value is nearly approximately one. So the result is that these variable are good predictor of the lighting over India.

\section{References}

[1] Litta A. J, Sumam Mary Idicula and C. Naveen Francis 2012. Artificial Neural Network Model for the Prediction of Thunderstorms over Kolkata. International Journal of ComputerApplications (0975 – 8887). Volume 50 – No.11, July 2012.

[2] Omvir Singh*, and Jagdeep Singh, Lightning fatalities over India: 1979–2011. Meteorol. Appl. 22: 770–778 (2015).

[3] Hugh J. Christian 1 , Richard J. Blakeslee 1 , Dennis J. Boccippio 1 , William L. Boeck 2, Dennis E. Buechler 3, Kevin T. Driscoll 3 , Steven J. Goodman 1 , 1John M. Hall 4 ,William J. Koshak 1 , Douglas M. Mach 3 , and Michael F. Stewart 3 , 2003. Global frequency and distribution of lightning as observed from space by the Optical Transient Detector, JOURNAL OF GEOPHYSICAL RESEARCH, VOL. 108, NO. D1, 4005, doi:10.1029/2002JD002347, 2003.

[4] D.M. Lal, S.D. Pawar, 2008. Relationship between rainfall and lightning over central Indian region in monsoon and premonsoon seasons, Atmospheric Research 92 (2009) 402–410.

[5] P. Murugavel, S. D. Pawar and V. Gopalakrishan*, Climatology of lightning over Indian region and its relationship with convective available potential energy. Int. J. Climatol. 34: 3179–3187 (2014).

[6] Philippe Lopez, Research Department, A lightning parameterization for the ECMWF model,
U.Schumann and H. Huntrieser, The global lightning-induced nitrogen oxides source, Atmos. Chem. Phys., 7, 3823–3907, 2007.

[7] D. M. Lal 1 , · Sachin D. Ghude 1 · M. Mahakur 1 · R. T. Waghmare 1 · S. Tiwari 1 Manoj K. Srivastava 2 · G. S. Meena 1. D. M. Chate 1 , 2000. Relationship between aerosol and lightning over Indo-Gangetic Plain (IGP), India, Clim Dyn (2018) 50:3865–3884.

[8] Daniel J. Cecil* , Dennis E. Buechler a , 2014, Richard J. Blakeslee b , Gridded lightning climatology from TRMM-LIS and OTD, Atmospheric Research 135–136 (2014) 404–414.

[9] Chen Lv, Member, IEEE, Yang Xing, Junzhi Zhang, Xiaoxiang Na, Yutong Li, Teng Liu,Dongpu Cao, Member, IEEE and Fei-Yue Wang, Fellow, IEEE, Levenberg-Marquardt Backpropagation Training of Multilayer Neural Networks for State Estimation of A Safety Critical Cyber-Physical System, 1551-3203 (c) 2017 IEEE.

[10] Devendraa Siingh a , ⁎ , P.S. Buchunde a , R.P. Singh b , Asha Nath a , Sarvan Kumar b , R.N. Ghodpage c , Lightning and convective rain study in different parts of India, Atmospheric Research 137 (2014) 35–48.

\end{document}